\documentclass[showpacs,twocolumn,showkeys,amsmath,amssymb,pra]{revtex4-1}

\usepackage{graphicx}   

\usepackage{color}

\begin{document}

\title{Entangling two distinguishable quantum bright solitons via collisions} 

\author{Thomas P.\ Billam}
\affiliation{Jack Dodd Center for Quantum Technology, Department of Physics, University of Otago, Dunedin 9016, New Zealand}

\author{Caroline L.\ Blackley}
\affiliation{Joint Quantum Centre  (JQC) Durham--Newcastle, Department of Chemistry, Durham University, Durham DH1 3LE, United Kingdom}

\author{Bettina Gertjerenken}
\affiliation{Institut f\"ur Physik, Carl von Ossietzky Universit\"at, D-26111 Oldenburg, Germany}

\author{Simon L.\ Cornish}
\author{Christoph Weiss}
\email{Christoph.Weiss@durham.ac.uk}
\affiliation{Joint Quantum Centre  (JQC) Durham--Newcastle, Department of Physics, Durham University, Durham DH1 3LE, United Kingdom}



\date{\today}
 
\begin{abstract}
{The generation of mesoscopic Bell states via collisions of distinguishable bright solitons has been suggested in Phys.\ Rev.\ Lett.\ {\bf 111}, 100406 (2013). Here, we extend our former proposal to two hyperfine states of \textsuperscript{85}Rb instead of two different atomic species, thus simplifying possible experimental realisations. A calculation of the $s$-wave scattering lengths for the hyperfine states $(f,m_f)=(2,+2)$ and $(3,+2)$ identifies parameter regimes suitable for the creation of Bell states with an advantageously broad Feshbach resonance. We show the generation of Bell states using the truncated Wigner method for the soliton's centre of mass and demonstrate the validity of this approach by a comparison to a mathematically rigorous effective potential treatment of the quantum many-particle problem.}
\end{abstract} 

\maketitle 


\section{\label{sec:intr}Introduction}

Bright solitons are a promising candidate to generate quantum entanglement for a mesoscopic number of atoms. Such bright solitons are realised experimentally in Bose-Einstein condensates~\cite{KhaykovichEtAl2002,StreckerEtAl2002,EiermannEtAl2004,CornishEtAl2006,Hulet2010b,MarchantEtAl2013}. These experiments have thus far been modelled by a mean-field description. However, going to lower particle numbers naturally requires a fuller quantum mechanical treatment.
The \emph{quantum bright solitons} described by such a treatment provide an excellent model system with which to investigate the ``middle-ground'' between quantum
and classical physics~\cite{HackermullerEtAl2004,Zurek2003}.

Scattering bright solitons off a single barrier was recently investigated in~\cite{ErnstBrand2010,WangEtAl2012,DamgaardHansenEtAl2012IOP,MartinRuostekoski2012,HelmEtAl2012,CuevasEtAl2013,AlvarezEtAl2013,PoloAhufinger2013IOP,FogartyEtAl2013,Gertjerenken2013} and references therein; with two barriers a soliton diode was suggested in~\cite{AsaduzzamanKhawaja2013}. In the regime of very low kinetic energies~\cite{WeissCastin2009,StreltsovEtAl2009b,GertjerenkenEtAl2012}, scattering a quantum bright soliton~\cite{CarterEtAl1987,LaiHaus1989,DrummondEtAl1993,CastinHerzog2001,CarrBrand2004,DelandeEtAl2013} off a barrier can even lead to Schr\"odinger cat states~\cite{WeissCastin2009,StreltsovEtAl2009b} that can be detected using their interference properties~\cite{WeissCastin2009,Gertjerenken2013}.

Schr\"odinger-cat states are highly non-classical superpositions\footnote{In a measurement, all particles would be on one side of the barrier; before the measurement, they were in a quantum superposition of all being on the right and all being on the left.} which are relevant for quantum-enhanced interferometry~\cite{GiovannettiEtAl2004}. The focus of our paper are mesoscopic Bell states
\begin{equation}
\label{eq:Bell}
|\psi_{\rm Bell}\rangle\equiv\frac1{\sqrt{2}}\left(|{\rm A},{\rm B}\rangle+e^{i\alpha}|{\rm B},{\rm A}\rangle\right),
\end{equation}
where $|A,B\rangle$ ($|B,A\rangle$)  signifies that the BEC A is on the left
(right) and the BEC B is on the right (left). 
While it might sound tempting to realise such mesoscopic quantum superpositions as, say, the ground states of Bose-Einstein condensate in a double well with carefully chosen signs and strengths of interactions, such an approach will not be successful in the presence of tiny asymmetries~(cf.~\cite{WeissTeichmann2007}) and decoherence. Suggestions of how such a state can be realised dynamically for Bose-Einstein condensates can be found in Refs.~\cite{MicheliEtAl2003,MahmudEtAl2005,WeissTeichmann2007,DagninoEtAl2009,GarciaMarchEtAl2011,MazzarellaEtAl2011} and references therein.

 Rather than using a potential to generate mesoscopic entanglement~\cite{WeissCastin2009,StreltsovEtAl2009b}, we have suggested to scatter two \textit{distinguishable}\/ quantum bright solitons off each other to generate mesoscopic Bell states~\cite{GertjerenkenEtAl2013}. Two colliding {distinguishable} bright solitons behave very differently from two colliding but initially indistinguishable solitons~\cite{LewensteinMalomed2009,HoldawayEtAl2013}: for indistinguishable solitons, either higher order nonlinear terms~\cite{LewensteinMalomed2009} (cf.~\cite{KhaykovichMalomed2006}) or additional harmonic confinement~\cite{HoldawayEtAl2013} are needed to generate entanglement. 
Quantum bright solitons have also been discussed in the context of symmetry breaking states~\cite{CastinHerzog2001}; for more general treatment of symmetry breaking in finite quantum systems see~\cite{BirmanEtAl2013} and references therein.

In this paper we discuss the generation of a mesoscopic Bell state via scattering two distinguishable bright solitons. While our original proposal~\cite{GertjerenkenEtAl2013} scattered two solitons of different species (\textsuperscript{85}Rb and \textsuperscript{133}Cs), we now suggest to use two hyperfine states of \textsuperscript{85}Rb. {This allows the generation of mesoscopic Bell states closer to the case of many photons which is an area of current theoretical and experimental research~\cite{StobinskaPRA2012,IskahovPRL2012}.} In addition to their inherent fundamental interest, such
states are of potential application as a resource in quantum information~\cite{IskahovPRL2012}.

Our paper is organised as follows: We first introduce the many-particle quantum model used to describe the two colliding solitons in sec.~\ref{sec:mode} before justifying our use of a classical field approach to describe mesoscopic quantum superpositions in sec.~\ref{sec:just}. In sec.~\ref{sec:suit} we describe a new Feshbach
resonance, offering excellent control over distinguishable soliton collisions, which we use for our numerics in sec.~\ref{sec:trunc}. {In sec.~\ref{sec:dist} we present signatures that distinguish quantum superpositions from statistical mixtures.} The paper ends with the conclusions in sec.~\ref{sec:conc}.

\begin{figure*}[ht]
\centering
\includegraphics[width=0.7\linewidth]{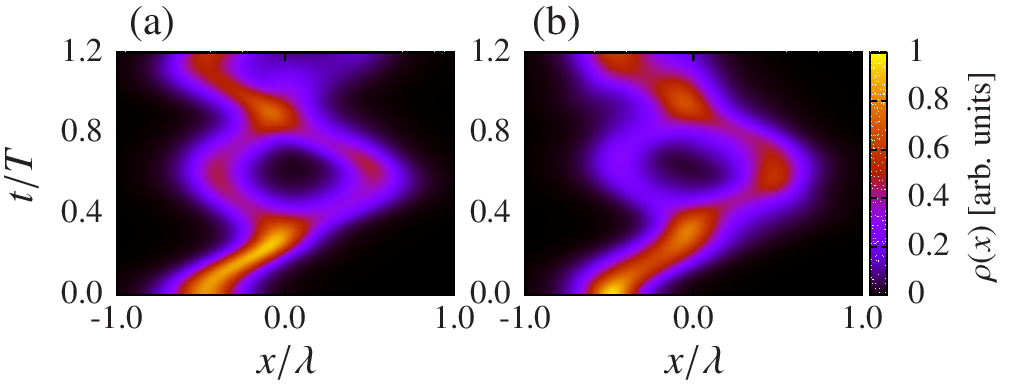}
	\caption{\label{fig:TruncWig} (a) Single-particle density for a
$N$-particle quantum bright soliton (soliton A) hitting a narrow, heavy
non-moving soliton (soliton B), computed using the effective potential
approach, as in ref.~\cite{GertjerenkenEtAl2013}.  (b) GPE simulation, using
centre-of-mass TW technique, of a single $N$-particle quantum bright soliton
colliding with the same single-particle potential due to soliton B as in the
effective potential treatment. Taking $m_{\rm A}=m_{\rm B}=m$ and $N_{\rm
B} g_{\rm B} = 10 N_{\rm A} g_{\rm A}$, the system can be described in terms of
the harmonic oscillator length $\lambda\equiv\lambda_{\rm A}$; we choose parameters such
that the mean initial displacement of the soliton  $-0.48\lambda$ and the
single-particle potential $V(x) = A\hbar \omega{\rm sech}^2(3x/2\lambda)$ with
$A \approx 1.2$ \cite{GertjerenkenEtAl2013}. $N_{\rm A}=100$. TW results averaged over 1000 realisations.}
\end{figure*}

\section{\label{sec:mode}Model}

In order to model two distinguishable solitons on the many-particle quantum level, we use the same approach as~\cite{GertjerenkenEtAl2013} and set $m_{\rm A}= m_{\rm B}$ at the end, where  $m_{\rm A}$ ($m_{\rm B}$) is the atomic mass of species A (B) (as we have two hyperfine states of the same species).
For our quasi-1D system, we consider an 
experimentally motivated harmonic confinement $\omega = 2\pi f$.
Mixtures of ultracold gases can be confined in a common optical trap with the same trap frequencies~\cite{SafronovaEtAl2006}, yielding
\begin{equation}
\omega = \frac{2\pi}T\;;\quad
\lambda_{\rm A} = \sqrt{\frac{\hbar}{m_{\rm A}\omega}}\;;\quad\lambda_{\rm B} = \sqrt{\frac{\hbar}{m_{\rm B}\omega}},
\end{equation}
where $\lambda_{\rm A}$ and $\lambda_{\rm B}$ are the harmonic oscillator lengths of the two species;
the interactions~$g_S=h f_{\perp} a_S$ are set by the scattering lengths~$a_S$ ($S=$ A,B or AB) and the perpendicular trapping-frequency, $f_{\perp}$~\cite{BergemanEtAl2003}.

We use the Lieb-Liniger model~\cite{LiebLiniger1963,SeiringerYin2008} for two species with additional harmonic confinement 
\begin{eqnarray}
\label{eq:H}
\hat{H} = &-&\sum_{j=1}^{N_{\rm A}}\frac{\hbar^2}{2m_{\rm A}}\partial_{x_j}^2 + \sum_{j=1}^{N_{\rm A}-1}\sum_{n=j+1}^{N_{\rm A}}g_{\rm A}\delta\left(x_j-x_{n}\right) \nonumber\\
&-&
\sum_{j=1}^{N_{\rm B}}\frac{\hbar^2}{2m_{\rm B}}\partial_{y_j}^2 + \sum_{j=1}^{N_{\rm B}-1}\sum_{n=j+1}^{N_{\rm B}}g_{\rm B}\delta\left(y_j-y_{n}\right) \nonumber\\
&+& \sum_{j=1}^{N_{\rm A}}\sum_{n=1}^{N_{\rm B}}g_{\rm AB}\delta\left(x_j-y_{n}\right) \nonumber\\
&+&
\sum_{j=1}^{N_{\rm A}}\frac 12 m_{\rm A}\omega^2x_j^2+\sum_{j=1}^{N_{\rm B}}\frac 12 m_{\rm B}\omega^2y_j^2\,,
\end{eqnarray}
where $x_j$ ($y_j$) and $g_{\rm A}<0$ ($g_{\rm B}<0$) are the atomic coordinates and intra-species interactions of species A (B), and 
 $g_{\rm AB}\ge 0$ is the inter-species interaction. 

We suggest to prepare the two solitons independently; for weak harmonic confinement a single soliton has the ground state energy (cf.~\cite{McGuire1964})
\begin{equation}
E_{\rm S}(N_{\rm S}) = -\frac{1}{24}\frac{m_{\rm S}g_{\rm S}^2}{\hbar^2}N_{\rm S}(N_{\rm S}^2-1)\,;\quad {\rm S} \in \{\rm A,B\}\,.
\end{equation}

Thus, our system has the total ground-state energy
\begin{equation}
E_0 = E_{\rm A}(N_{\rm A}) + E_{\rm B}(N_{\rm B}) \,.
\end{equation}
The total kinetic energy related to the centre-of-mass momenta $\hbar K_{\rm S}$ ( ${\rm S}  \in \{\rm A,B\}$) of the two solitons reads 
\begin{equation}
E_{\rm kin} = \frac{\hbar^2 K_{\rm A}^2}{2N_{\rm A}m_{\rm A}} + \frac{\hbar^2 K_{\rm B}^2}{2N_{\rm B}m_{\rm B}}.
\end{equation}

We extend the low-energy regime investigated for single-species solitons in Refs.~\cite{WeissCastin2009,SachaEtAl2009,GertjerenkenEtAl2012} to two species: 
\begin{equation}
\nonumber
E_{\rm kin} < \min\{\Delta_{\rm A}, \Delta_{\rm B}\},\quad
\Delta_{\rm S} =  \left|E_{\rm S}(N_{\rm S}-1)- E_{\rm S}(N_{\rm S})\right|.
\end{equation}
In this energy regime, each of the quantum matter-wave bright solitons is energetically forbidden to break up into two or more parts.
Highly entangled states are characterised by a roughly 50:50 chance of finding the soliton~A (B) on the left/right combined with a left/right correlation close to one indicating that whenever soliton~A is on the one side, soliton~B is on the other:
\begin{multline}
\label{eq:corr}
\gamma(\delta)\equiv
\int_{\delta}^{\infty}dx_1\ldots\int_{\delta}^{\infty}dx_{N_{\rm A}}\int^{-\delta}_{-\infty}dy_1\ldots\int_{-\infty}^{-\delta}dy_{N_{\rm B}}|\Psi|^2\\
+\int^{-\delta}_{-\infty} dx_1\ldots\int^{-\delta}_{-\infty}dx_{N_{\rm A}}\int_{\delta}^{\infty}dy_1\ldots\int_{\delta}^{\infty}dy_{N_{\rm B}}|\Psi|^2\,,
\end{multline}
where $\Psi= \Psi(x_1,\ldots,x_{N_{\rm A}},y_1,\ldots,y_{N_{\rm B}})$ is the many-particle wave function (normalised to one) and $\delta \ge 0$.  The correlation $\gamma(\delta)$ will serve as an indication of entanglement: Bell states (\ref{eq:Bell}) are characterised by $\gamma\simeq 1$ combined with a 50:50 chance to find soliton A either on one side or on the other.

Behaviour for larger particle numbers can be described by the Gross-Pitaevskii equation (GPE) (cf.~\cite{PuBigelow1998,Timmermans1998,OehbergSantos2001,HeEtAl2012})
\begin{align}
i\hbar\partial_t\varphi_{\rm A}(x,t) =& \left[-\frac{\hbar^2}{2m_{\rm A}}\partial_x^2+\frac{g_{\rm A}}2|\varphi_{\rm A}(x,t)|^2\right]\varphi_{\rm A}(x,t)\nonumber\\
&
+\left[\frac12m_{\rm A}\omega^2x^2 +\frac{g_{\rm AB}}2|\varphi_{\rm B}(x,t)|^2\right]\varphi_{\rm A}(x,t)\phantom{,}\nonumber\\
i\hbar\partial_t\varphi_{\rm B}(x,t) =& \left[-\frac{\hbar^2}{2m_{\rm B}}\partial_x^2+\frac{g_{\rm B}}2|\varphi_{\rm B}(x,t)|^2\right]\varphi_{\rm B}(x,t)\nonumber\\
&
+\left[\frac12m_{\rm B}\omega^2x^2 +\frac{g_{\rm AB}}2|\varphi_{\rm A}(x,t)|^2\right]\varphi_{\rm B}(x,t)\;,\nonumber
\end{align}
where the single-particle density $|\varphi_{\rm S}(x,t)|^2$ is normalised to $N_{\rm S}$ ($ {\rm S} \in \{\rm A,B\}$).

\section{\label{sec:just}Justifying Truncated Wigner for the centre of mass}

When hitting a barrier, the generic behaviour of a mean-field bright soliton is
to break into two parts; the fraction of the atoms transmitted decreases for
increasing potential strength (cf.~\cite{MartinRuostekoski2012,
CuevasEtAl2013}). An analogous behaviour also occurs when two distinguishable
mean-field bright solitons collide with each other, as shown in the Supplemental
Material of~\cite{GertjerenkenEtAl2013}. Only at very low kinetic
energies~\cite{WeissCastin2009,GertjerenkenEtAl2012,GertjerenkenEtAl2013} do
mesoscopic quantum superpositions occur as a result of such collisions.

To describe low kinetic energy collisions of two distinguishable bright
solitons, taking into account the formation of mesoscopic quantum
superpositions, we combine mean-field calculations via the GPE with
Truncated-Wigner Approximation (TWA) for the
centre of mass degree of freedom in order to model true quantum
behaviour~\cite{GertjerenkenEtAl2013}. {The truncated-Wigner approximation
(TWA) describes quantum systems by averaging over realisations of an
appropriate classical field equation (in this case,
the GPE) with initial noise
appropriate to either finite~\cite{BieniasEtAl2011} or zero
temperatures~\cite{MartinRuostekoski2012}. While the GPE assumes both position
and momentum are well-defined, this is not true for a single quantum particle
of finite mass for which, in general, both position and momentum involve
quantum noise satisfying the uncertainty relation. Our TWA calculations for the
soliton centre-of-mass wave function use Gaussian probability distributions for
both (satisfying minimal uncertainty)~\cite{GertjerenkenEtAl2013}.}

This centre-of-mass TW technique can be justified by comparison to the
rigorously proved~\cite{WeissCastin2012} effective potential
approach~\cite{SachaEtAl2009,WeissCastin2009}: In fig.~\ref{fig:TruncWig} we
compare the single-particle effective potential treatment
[fig.~\ref{fig:TruncWig}(a)] for the case of a low-mass bright soliton
colliding with a heavy bright soliton with a centre-of-mass TW GPE simulation
[fig.~\ref{fig:TruncWig}(b)] using the same effective single-particle
potential. In the low kinetic energy regime considered, the low-mass bright
soliton is either completely reflected or completely transmitted in any
individual realisation. The good level of agreement up to the time where the
solitons re-collide confirms that the centre-of-mass TW technique can
successfully capture the dynamical formation of quantum superpositions in the
centre-of-mass coordinate, as required.

\section{\label{sec:suit}Suitable Feshbach resonance}

\begin{figure}[ht]
\includegraphics[width=0.97\linewidth]{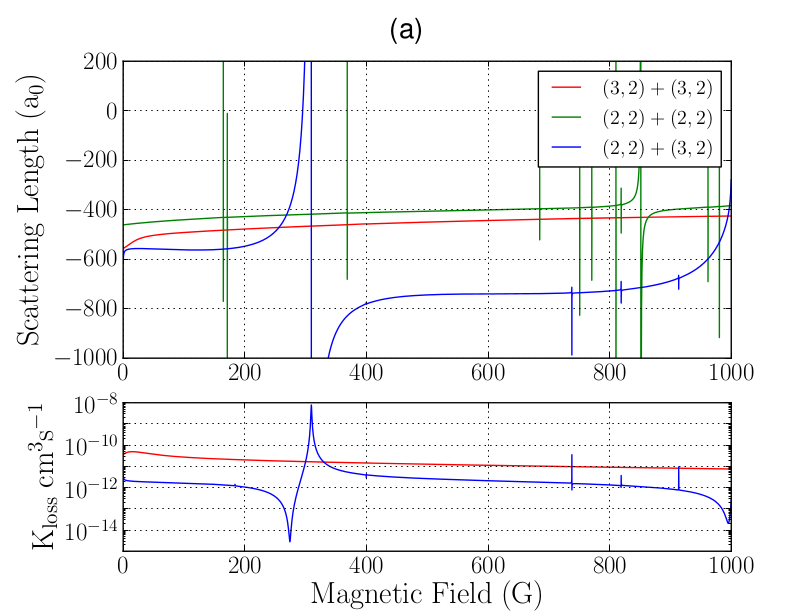}\\
\includegraphics[width=0.97\linewidth]{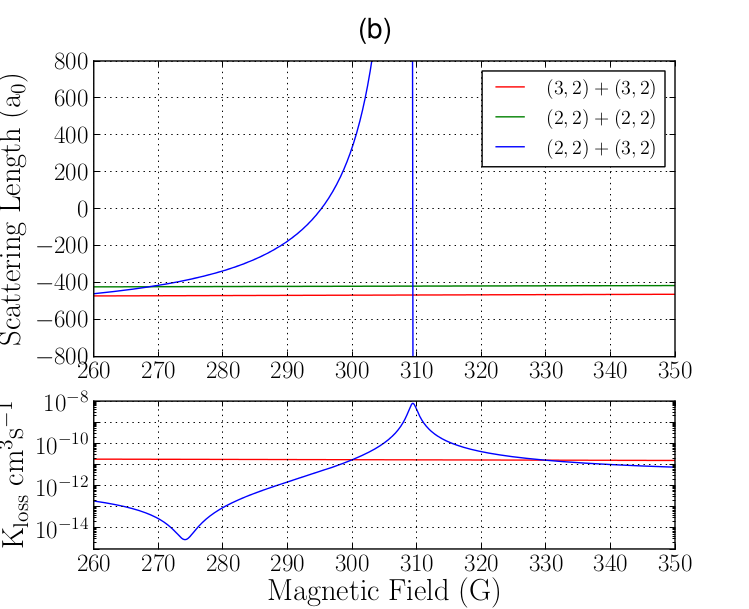}
\caption{The $s$-wave scattering lengths for the $(f,m_{\rm f}$)= (2,+2), (3,+2) and (2,+2)+(3,+2) states of $^{85}$Rb. (a) The scattering length is split into real and imaginary components, the real part is shown in the top plot, the imaginary part is proportional to the inelastic decay rate-coefficent $K_{\rm loss}$, shown in the lower graph. (b) Zoom of (a), the wide resonance in the mixed spin state allows for tuning of the scattering length.
}
\label{fig:feshbach_resonance}
\end{figure}
Using mixed states of the same atomic species allows for the creation of distinguishable solitons while removing the need for a dual-species laser cooling apparatus. 
The physical requirements for the experiment are a negative background scattering length for each of the two distinguishable soliton states, and a wide Feshbach resonance in the mixed-state scattering length.

Coupled-channels calculations were performed as detailed in Ref \cite{BlackleyEtAl2013} on each of the $(f_a,f_b)=(2,3)$ hyperfine manifold of $^{85}$Rb$_2$, using the {\sc molscat} program \cite{MOLSCAT} adapted to handle collisions in external fields \cite{GonzalezMartinezEtAl2007}. A wide tunable resonance was found in the $(f_a,m_{f_a})(f_b,m_{f_b}) = (2,2)(3,2)$ channel. The resonance has a width of $\Delta$=14~G determined by the difference between the zero-crossing and the pole in the scattering length. Whilst excited-state resonances are subject to decay from inelastic collisions \cite{Hutson:res:2007} the resonance has $a_{\rm res} > 10,000~a_0$ making it `pole-like' from an experimental point of view. In the excited states the complex scattering length is given by $a(B)=\alpha(B) - i\beta(B)$, where $\alpha(B)$ is the real part of the scattering length, and $\beta(B)$ the imaginary part of the scattering length is proportional to the rate-coefficient for 2-body losses due to inelastic collisions, $K_{\rm loss}=\frac{2h}{\mu}g_n\beta(B)$, where $g_n$=1 (2) for a BEC of distinguishable (indistinguishable) particles. The real part of the scattering length and associated plots of $K_{\rm loss}$, of both the mixed-state and the individual states, are shown in fig.~\ref{fig:feshbach_resonance}. Note that $K_{\rm loss}=0$ for the absolute internal ground state $(f,m_f) = (2,+2)$.

\begin{figure*}[ht]
\centering
\includegraphics[width=0.7\textwidth]{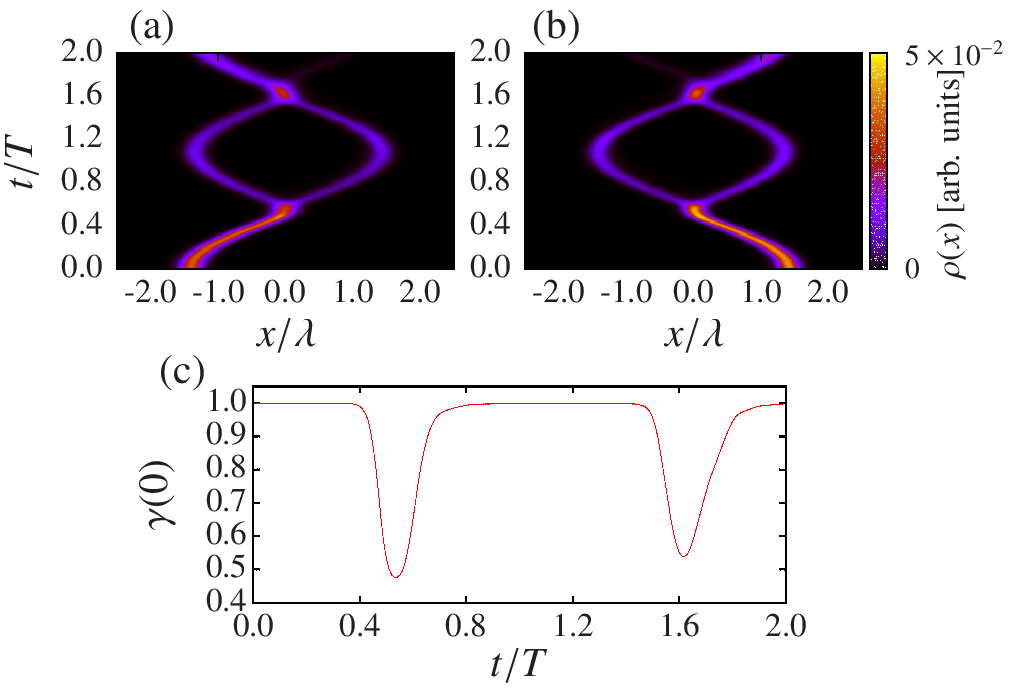}
\caption{\label{fig:TWA2Component} Centre-of-mass TW GPE simulation of a
two-component collision of solitons in the $(f,m_f)=(2,+2)$ and $(3,+2)$
hyperfine states of $ ^{85}$Rb. Parameters are $a_{(2,2)}=-410 a_0$,
$a_{(3,2)}=-460 a_0$, $N_{(2,2)} = N_{(3,2)} \approx 90$, $f
= 2\,{\rm Hz}$, $f_\perp = 50\, {\rm Hz}$, and $a_{(2,2) / (3,2)} \simeq
30.0 a_0$ (conveniently reached at around 295\,G, see fig.~\ref{fig:feshbach_resonance}).
The initial displacement of the solitons is $\approx \pm 10.1 \mu{\rm m}$.
Panels (a), (b) and (c) respectively show the average single-particle densities
of the $(2,2)$ and $(3,2)$ components, and the left/right
correlation $\gamma(0)$. 1000 realisations were performed.}
\end{figure*}

The three-dimensional scattering calculations can be converted into a one dimensional interaction parameter $g$ by taking account of the trapping frequency ($f_{\perp}$). With the introduction of the trapping parameters it is possible to cause a confinement induced resonance (CIR) as predicted in \cite{Olshanii1998} when $a_{\perp}\approx C a_{\rm 3D}$. However, given the confinement parameters for this problem ($f_\perp = 50\,{\rm Hz}$ and $f = 2\,{\rm Hz}$, see fig.~\ref{fig:TWA2Component}), the CIR would occur when $a_{\rm 3D} \approx 3.5\times 10^5 a_0$ which would not interfere with any practical implementation.

\section{\label{sec:trunc}Truncated Wigner for the centre of mass for two distinguishable bright solitons}
Using the Feshbach resonance described in the previous section we perform a
centre-of-mass TW GPE simulation for the two-component GPE using parameters for a mixture of the $(f,m_f)=(2,+2)$ and $(3,+2)$ hyperfine states of $^{85}$Rb.
 The resulting average density profiles for
the two components, and the left/right correlation $\gamma(0)$ are shown in
fig.~\ref{fig:TWA2Component}. The high ($\approx 1$) value of $\gamma(0)$
subsequent to the first collision indicates the formation of a Bell state with
high fidelity. Compared to the $^{85}$Rb -- $^{133}$Cs scheme suggested in
ref.~\cite{GertjerenkenEtAl2013}, the present scheme is feasible at higher atom
numbers, less sensitive to magnetic bias field strength, and generates
higher-fidelity Bell states. These factors make the present scheme an even more
experimentally attractive proposal to generate Bell states of distinguishable
bright solitons.

\section{\label{sec:dist}Distinguishing quantum superpositions from statistical mixtures}

Bell inequalities, which are both interesting because they allow to fundamentally test our understanding of quantum mechanics~\cite{ClauserShimony1978,AspectEtAl1982} and because of their importance for quantum cryptography~\cite{Ekert1991}, are still a topic of current research~\cite{TorlaiEtAl2013}. For mesoscopic Bell states, related separability conditions are available{~\cite{SimonBouwmeester03,IskahovPRL2012}. For a bipartite photonic system a violation of the inequality
\begin{equation}\label{eq:sepcond}
\sum_{k=1}^3\Delta S_k^2/\langle  S_0 \rangle \ge 2
\end{equation}
has been shown to be a sufficient condition of non-separability and has been used to identify polarisation entanglement for squeezed vacuum pulses~\cite{IskahovPRL2012}. Here, $S_k = S_k^A + S_k^B$ denote the Stokes parameters~\cite{SimonBouwmeester03} and $\langle  S_0 \rangle$ is the total photon number.  To convey condition~({\ref{eq:sepcond}}) to our situation the properties left and right would take on the role of horizontal and vertical polarisation.

In addition to the above, in the collisions we consider here the
interference properties discussed in  \cite{GertjerenkenEtAl2013} for two different species would also
be available to distinguish quantum superpositions from statistical mixtures.

\section{\label{sec:conc}Conclusion}

We have investigated numerically the generation of mesoscopic Bell states via the collision of two distinguishable quantum bright solitons. For experimentally realistic parameters, we have used Truncated Wigner for the centre of mass~\cite{GertjerenkenEtAl2013} (which we justified further) to predict entanglement generation.  We have in particular extended the scheme suggested in~\cite{GertjerenkenEtAl2013} for two bright solitons of two different species to two solitons of two distinct hyperfine states of the same species, providing several advantages compared to the original suggestion~\cite{GertjerenkenEtAl2013}:

\begin{enumerate}
\item We predict a much broader Feshbach resonance (fig.~\ref{fig:feshbach_resonance}~b) then for the two-species case investigated in~\cite{GertjerenkenEtAl2013}. This will considerably simplify future experiments.
\item We predict a higher left/right correlation in the Bell state (fig.~\ref{fig:TWA2Component}~c),
potentially aiding experimental detection.
\item Only a Bose-Einstein condensate of one species is required; the two
  distinguishable bright solitons could be produced from a single initial
  Bose-Einstein condensate.
\item The current situation is closer to the mesoscopic Bell states for photons of refs.~\cite{StobinskaPRA2012,IskahovPRL2012}.
\end{enumerate}

\acknowledgments
We thank S.\ A.\ Gardiner, J.\ L.\ Helm, J.\ M.\ Hutson, C.\ R.\ Le~Sueur, L.\ Khaykovich for discussions.
We thank the Marsden Fund
of New Zealand (Contract No. UOO162) and the Royal
Society of New Zealand (Contract No. UOO004) (T. P. B.),
the Faculty of Science at Durham University (C. L. B.),  and the UK EPSRC (Grant No. EP/G056781/1 and EP/K03250X/1) (C. W.) for
funding.

\section*{References}
%

\end{document}